\documentclass[pre,twocolumn,amsmath,amssymb,showpacs]{revtex4}
\usepackage{graphicx}
\usepackage{dcolumn}
\usepackage{bm}
\begin{document}
\newcommand{\widthfigA}{0.45\textwidth}
\newcommand{\bfGamma}{\mbox{\boldmath $\bf\Gamma$}}
\newcommand{\condiIaa}{[I]}
\newcommand{\condiIIa}{[I$\!$IA]}
\newcommand{\condiIIb}{[I$\!$IB]}
\newcommand{\condiIII}{[I$\!$I$\!$I]}
%
%
%
%
\title{Critical bending point in the Lyapunov localization spectra 
of many-particle systems
}   
\author{Tooru Taniguchi and Gary P. Morriss}
\affiliation{School of Physics, University of New South Wales, 
Sydney, New South Wales 2052, Australia}
\date{\today}
\begin{abstract}
   The localization spectra of Lyapunov vectors in many-particle 
systems at low density exhibit a characteristic bending behavior. 
   It is shown that this behavior is due to a restriction on the maximum 
number of the most localized Lyapunov vectors determined by 
the system configuration and mutual orthogonality. 
   For a quasi-one-dimensional system this leads to a predicted 
bending point at $n_{c} \approx 0.432 N$ for an $N$ particle system. 
   Numerical evidence is presented that confirms this predicted 
bending point as a function of the number of particles $N$. 
\end{abstract}
%
\vspace{1cm}
\maketitle
%
%
%
%
 
   The Lyapunov spectrum is an indicator of dynamical instability 
in the phase space of many-particle systems. 
   It is introduced as the sorted set $\{\lambda^{(n)}\}_{n}$, 
$\lambda^{(1)}\geq \lambda^{(2)}\geq \cdots$ of Lyapunov exponents 
$\lambda^{(n)}$,  which give the exponential rates of expansion 
or contraction of the distance between 
nearby trajectories (Lyapunov vector) 
and is defined for each independent component of the phase space. 
   In Hamiltonian systems and some thermostated systems, a 
symmetric structure of the Lyapunov spectra, the so called 
the conjugate pairing rule, is observed \cite{Dre88,Eva90,Det96a,Tan02a}.  
   One of the most significant points of the Lyapunov spectrum is that 
each Lyapunov exponent indicates a time scale given 
by the inverse of the Lyapunov exponent so we can consider the Lyapunov 
spectrum as a spectrum of time-scales. 
   The smallest positive Lyapunov exponent region of the spectrum 
is dominated by macroscopic time and length scale
behavior, and  here some delocalized mode-like structures (the Lyapunov 
modes) have been observed in the Lyapunov vectors 
\cite{Pos00,Eck00,Tan02c,Tan03a,Mar04,Wij03,Yan04,Tan04a,For04,Tan04b}. 
   On the other hand, the largest Lyapunov exponent region of 
Lyapunov spectrum is dominated by short time scale behavior, 
and in this region the Lyapunov vectors are localized 
(Lyapunov localization). 
   The position of the localized region of 
Lyapunov vectors moves as a function of time. 
   A variety of many-particle systems show Lyapunov localization, 
for example, the Kuramoto-Sivashinsky model \cite{Man85}, 
a random matrix model \cite{Liv89}, 
map systems \cite{Kan86,Fal91,Gia91,Pik98}, 
coupled nonlinear oscillators \cite{Pik01}, 
and many-disk systems \cite{Mil02,Tan03b}.

%

   Recently, a quantity to measure  
strength of Lyapunov localization was proposed \cite{Tan03b}.
   For each Lyapunov exponent we construct the normalized 
Lyapunov vector component amplitude 
$\gamma_{j}^{(n)}\equiv |\delta\bfGamma_{j}^{(n)}|
/|\delta\bfGamma^{(n)}|$ for each particle $j$.
   Here $\delta\bfGamma^{(n)} \equiv 
(\delta\bfGamma_{1}^{(n)},\delta\bfGamma_{2}^{(n)},
\cdots,\delta\bfGamma_{N}^{(n)})$ is the Lyapunov vector 
for the $n$-th Lyapunov exponent $\lambda^{(n)}$, 
and $\delta\bfGamma_{j}^{(n)}$ is the contribution of the 
$j$-th particle to the $n$-th Lyapunov vector.  
      The localization of the $n$-th Lyapunov vector is then
\begin{eqnarray}
   \mathcal{W}^{(n)} \equiv 
   \exp\left[
      - \sum_{j=1}^{N} \left\langle
         \gamma_{j}^{(n)} \ln \gamma_{j}^{(n)} 
      \right\rangle
   \right] .
\label{LocalWidth}\end{eqnarray}
   The bracket $\langle\cdots\rangle$ in Eq. 
(\ref{LocalWidth}) indicates the time-average. 
   The quantity $- \sum_{j=1}^{N} \langle
\gamma_{j}^{(n)} \ln \gamma_{j}^{(n)}\rangle$ 
in the definition (\ref{LocalWidth}) of 
$\mathcal{W}^{(n)}$
can be regarded as an entropy-like 
quantity, as $\gamma_{j}^{(n)}$ is a distribution function 
over the particle index $j$.   
   The quantity $\mathcal{W}^{(n)}$ satisfies the inequality 
$1\leq \mathcal{W}^{(n)}\leq N$, and can be interpreted as the 
effective number of particles contributing to the non-zero 
components of the Lyapunov vector. 
   In a Hamiltonian system it satisfies the conjugate 
relation $\mathcal{W}^{(\mathcal{D}-j+1)}=\mathcal{W}^{(j)}$ for 
any $j$ with the phase space dimension $\mathcal{D}$, 
because of the symplectic structure. 

   The set of quantities $\{\mathcal{W}^{(n)}\}_{n}$ 
which we call the {\it Lyapunov localization spectrum}, 
has been calculated previously in many-particle 
systems with hard-core interactions \cite{Tan03b} and 
with soft-core interactions \cite{For04}. 
   In these systems, the value of $\mathcal{W}^{(n)}$ 
usually increases with the Lyapunov index $n$, and this implies  
that Lyapunov vectors for the largest exponent region 
are the most localized. 
   The quantity $\mathcal{W}^{(n)}$ has a minimum 
value of 2, as a minimum of two particles
are involved in each collision, and it has been shown numerically 
that the value of $\mathcal{W}^{(n)}$ for  
the largest Lyapunov exponent approaches 2 as 
the density approaches zero \cite{Tan03b}.  
   It was also shown that $\mathcal{W}^{(n)}$ can detect 
not only the localized behavior of Lyapunov vectors, but also 
the de-localized behavior observed in the Lyapunov modes \cite{Tan03b}.

%

\begin{figure}[t]
\begin{center}
\includegraphics[width=\widthfigA]{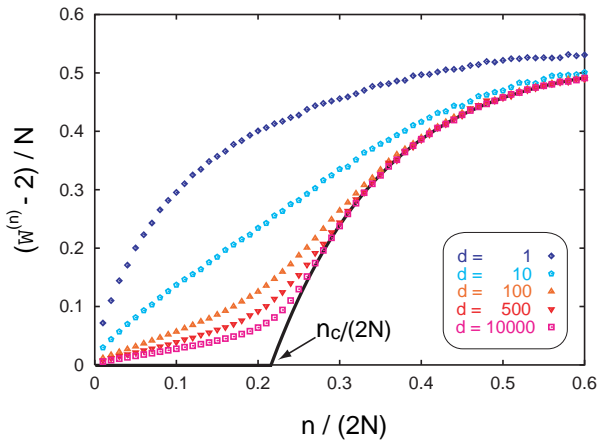}
\end{center}
\caption{ (Color online) 
      The normalized Lyapunov localization spectra 
   $\{(\mathcal{W}^{(n)}-2)/N\}_n$ as functions of the normalized 
   Lyapunov index $n/(2N)$ for quasi-one-dimensional $50$ hard-disk 
   systems at different densities given by $d=1$ (diamonds),  
   $d=10$ (circles), $d=100$ (triangles), 
   $d=500$ (inverted triangles), and $d=10000$ (squares). 
       The parameter $d$ is inversely proportional to the density 
   as $\rho = \pi R^{2}/[(1+d)L_{y}^{2}]$.   
      The solid line is the asymptotic form of Lyapunov 
   localization spectrum as the density approaches zero, and 
   $n\approx n_{c}$ is its bending point. 
   }
\label{fig1CriValDensi}\end{figure}  


   One of the important characteristics of the Lyapunov 
localization spectrum is its bending behavior at low density 
\cite{Tan03b}. 
   In Fig. \ref{fig1CriValDensi} we show an example of 
such bending behavior \cite{note1} in a quasi-one-dimensional system 
of $50$ hard disks with periodic boundary conditions. 
   Here the radius of the particles $R=1$, the mass of each particle $m=1$, 
the total energy of system $E=N$, and the system size  
$L_{y} = 2R(1+10^{-6})$ and  $L_{x} = NL_{y}(1+d)$ 
with parameter $d$, related to the density by 
$\rho = \pi R^{2}/[(1+d)L_{y}^{2}]$. 
   In this system the particles remain ordered because 
the vertical size $L_{y}$ prohibits the exchange of particles 
\cite{Tan03a,Tan03b,Tan04a}.  
   In Fig. \ref{fig1CriValDensi}, five different Lyapunov localization spectra  
are represented at different densities. 
   In the low density limit, this figure shows a bending point 
in the Lyapunov localization spectra at $n_{c}$.
   At this bending point $n\approx n_{c}$ the Lyapunov 
localization spectra $\mathcal{W}^{(n)}$ changes from a 
linear dependence upon 
the Lyapunov index ($n$) to an exponential dependence.
   This bending becomes sharper at lower density, and the numerical 
results suggest that in the low density limit 
the Lyapunov localization spectrum converges to the solid 
line given by 
\begin{eqnarray}
   \mathcal{W}^{(n)} = 
   \left\{
      \begin{array}{ll}
      2  & \mbox{in}\;\;\; n < n_{c} \\
      \gamma -  \alpha N \exp\{-\beta n/(2N)\}
      &\mbox{in} \;\;\; n \geq n_{c} 
      \end{array}
     \right. 
\label{LowDensityLimitLine}\end{eqnarray}
with constants $\alpha$, $\beta$, and $\gamma 
\equiv 2+\alpha N \exp\{-\beta n_{c}/(2N)\}$ 
[so that $\mathcal{W}^{(n)}|_{n=n_{c}}=2$ 
in Eq. (\ref{LowDensityLimitLine})]. 
   In other words, we can estimate 
the critical value $n\approx n_{c}$ defined 
as the bending point of the Lyapunov localization spectrum 
by the value of the fitting parameter 
$n_{c}$ in the fit of the Lyapunov localization spectrum 
to the function (\ref{LowDensityLimitLine}). 
 
   The bending behavior of Lyapunov localization spectra, 
shown in Fig. \ref{fig1CriValDensi} is associated with a similar 
bending point in the Lyapunov exponent spectra.  
   Moreover, the existence of a linear dependence of 
$\mathcal{W}^{(n)}$ on $n$ in the region $n \leq n_{c}$, 
at low density,  
is connected with some other known kinetic properties, 
for example, that the mean free time is inversely proportional 
to density, and the Krylov relation, that the 
largest Lyapunov exponent $\lambda^{(1)}$ depends 
on the density $\rho$ like $\lambda^{(1)} \sim -\rho \ln \rho$ 
\cite{Tan03b}.    
   These results suggest that the existence of the linear dependence of 
$\mathcal{W}^{(n)}$ is connected to the density range where
kinetic theory provides an accurate description.  
   These points were investigated in detail in 
quasi-one-dimensional systems, although 
the bending behavior of Lyapunov localization spectra is also 
observed in fully two-dimensional square systems \cite{Tan03b}. 
   However, no mechanism was proposed for this bending behavior.  

   In this paper we construct a mechanism 
that leads to the bending behavior observed in the Lyapunov localization spectra, 
and predicts the critical value $n_{c}$ for 
quasi-one-dimensional systems.


   We begin by recalling some properties of the Lyapunov vectors 
of many-particle systems. 
   The first property expresses the fact that different  
Lyapunov vectors sample different directions in phase space. 
%
\begin{description}
\item[\condiIaa] Lyapunov vectors with different Lyapunov indexes 
are orthogonal. 
\end{description}
%
   The second property of Lyapunov vectors is based on the fact 
that particle interactions occur between two different particles:
%
\begin{description}
\item[\condiIIa] In the low density limit, all the 
Lyapunov vectors $\delta\bfGamma^{(n)}$, for $n< n_{c}$ 
have non-zero components for only two particles.
\end{description}
%
   This leads to the known result that 
$\mathcal{W}^{(1)} \rightarrow 2$ as $\rho \rightarrow 0$. 
   The third property is justified only for quasi-one-dimensional
systems, and restricts the property \condiIIa$\,$ further to
%
\begin{description}
\item[\condiIIb]  Two particles, whose Lyapunov vector 
   components of $\delta\bfGamma^{(n)}$, $n< n_{c}$ 
   are non-zero values in the low density limit, 
   are nearest-neighbors. 
\end{description}
%
   In Ref. \cite{Tan03b} it is shown that in the largest Lyapunov 
exponent region, at low density, two 
non-zero particle components of the Lyapunov vector appear 
by particle collisions. 
   Moreover, in quasi-one-dimensional systems, 
particle collisions occur only between nearest neighbor particles. 
   Based on these facts, the property \condiIIb$\,$ is justified. 

   Using the properties \condiIIa$\,$ and 
\condiIIb, in the low density 
limit, the Lyapunov vectors $\delta\bfGamma^{(n)}$
in the largest Lyapunov exponent region can be represented as 
\begin{eqnarray}
 \delta\bfGamma^{(n)} \sim (
 \mathbf{0}, \mathbf{0}, \cdots,\mathbf{0}, 
 \delta\bfGamma_{\mu_{n}}^{(n)},
 \delta\bfGamma_{\mu_{n}+1}^{(n)}, \mathbf{0},
 \cdots,\mathbf{0}) 
\label{LyapuVectTwo}\end{eqnarray}
where $\mathbf{0}$ is the null vector. 
   Here, the particle numbering for 
the quasi-one-dimensional system is 
from left to right, and $\mu_{n}$ and 
$\mu_{n}+1$ are nearest neighbor particles 
whose Lyapunov vector components are non-zero. 
   (We put $\delta\bfGamma_{N+1}^{(n)}\equiv 
\delta\bfGamma_{1}^{(n)}$ for periodic boundary conditions 
in the longitudinal direction.)
   In general, the particle number $\mu_{n}$ depends 
on the Lyapunov index $n$ and on time. 
   Now we consider the restriction imposed by condition 
\condiIaa$\,$ on the form (\ref{LyapuVectTwo}) of 
Lyapunov vector. 
   To satisfy condition \condiIaa, the particle numbers 
$\mu_{n}$ and $\mu_{n}+1$ in the Lyapunov vector 
(\ref{LyapuVectTwo}) must be different for different 
Lyapunov indexes in $n < n_{c}$. 
   Here, we assume that the vector components for particle $j$,
non-zero components $\delta\bfGamma_{j}^{(n)}$ and 
$ \delta\bfGamma_{j}^{(n')}$ are themselves not 
orthogonal for different Lyapunov indices $n$ and $n'$. 
   This restricts the number of independent 
Lyapunov vectors of the form (\ref{LyapuVectTwo}), and puts
an upper limit on the 
critical value $n_{c}$.

\begin{figure}[t]
\begin{center}
\includegraphics[width=\widthfigA]{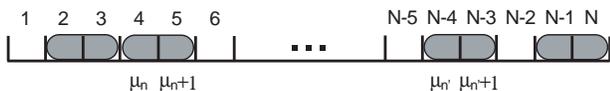}
\end{center}
\caption{
      A schematic illustration of the randomly distributed 
   brick model, used to explain the maximum number of independent 
   Lyapunov vectors with only two non-zero particle components. 
      The box labelled $j$ corresponds to 
   the Lyapunov vector component of the $j$-th particle 
   ($j=1,2,\cdots,N)$. 
      Gray-filled rectangular bricks occupying
   boxes $\mu_{n}$ and $\mu_{n}+1$ represent possible 
   non-zero components of 
   Lyapunov vectors for which $n <n_{c}$.  
   The value $n_{c}$ is the average number of bricks
   that can be randomly arranged on the set of boxes.
   }
\label{fig2BrickPack}\end{figure}  

   This situation is explained using a simple model 
whose schematic illustration is given in Fig. \ref{fig2BrickPack}. 
   In this  {\it randomly distributed 
brick model}, N boxes corresponding  
to each of the particles are arranged on a line. 
   Each Lyapunov vector represented by Eq. (\ref{LyapuVectTwo}) 
must have non-zero components for only two particles 
$\delta\bfGamma_{\mu_{n}}^{(n)}$ and 
$\delta\bfGamma_{\mu_{n}+1}^{(n)}$.
These are shown as 
a gray-filled rectangular brick filling 
boxes $\mu_{n}$ and $(\mu_{n}+1)$ 
in Fig. \ref{fig2BrickPack}. 
   To satisfy condition \condiIaa, these bricks 
must not overlap. 
   The critical value $n_{c}$ is then the average number 
of bricks which can be put without overlaps 
on the $N$ different boxes.

\begin{figure}[t]
\begin{center}
\includegraphics[width=\widthfigA]{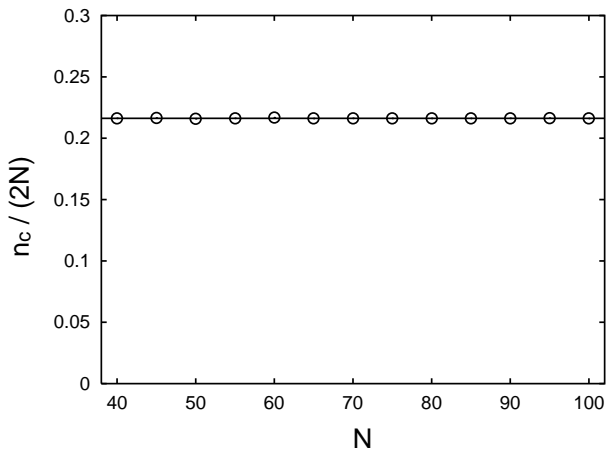}
\end{center}
\caption{
   The normalized critical value $n_{c}/(2N)$ given by 
   the randomly distributed brick model 
   as a function of the number of particles $N$.   
      The solid line is given by fitting the numerical data 
   to a constant function  $n_{c}/(2N) = \xi$ with 
   the fitting parameter $\xi \approx 0.216$. 
   }
\label{fig3BrickCriVal}\end{figure}  
   
   There is one more important point to get an explicit 
value of $n_{c}$ from the above mechanism: 
%
\begin{description}
\item[\condiIII] The particle number $\mu_{n}$ 
   in Eq. (\ref{LyapuVectTwo}) is chosen 
   randomly with respect to the Lyapunov index $n (< n_{c})$, 
   so that there is no overlap among the non-zero 
   Lyapunov vector components for Lyapunov vectors  
   with different Lyapunov indices. 
\end{description}
%
   This means that the bricks  
shown in Fig. \ref{fig2BrickPack} 
must be randomly distributed without overlaps.  
   Therefore randomly constructed 
configurations with any number of single empty boxes at 
non-neighboring positions are possible.  
   Taking an ensemble average of the values of $\tilde{n}_{c}$ 
for each possible configuration
we obtain the critical value $n_{c}$. 
   Figure \ref{fig3BrickCriVal} shows the normalized critical 
value $n_{c}/(2N)$ from such a numerical simulation 
for different numbers of particles $N$. 
   The data in this figure shows that $n_{c}/(2N)$ is independent 
of $N$, giving an estimate of the normalized  critical value as 
\begin{eqnarray}
   \frac{n_{c}}{2N} \approx  0.216. 
\label{CritiValue}\end{eqnarray}
%
   The accuracy of this value can be used as a check of the 
proposed mechanism for 
the bending behavior of Lyapunov localization 
spectra at low density.

\begin{figure}[t]
\begin{center}
\includegraphics[width=\widthfigA]{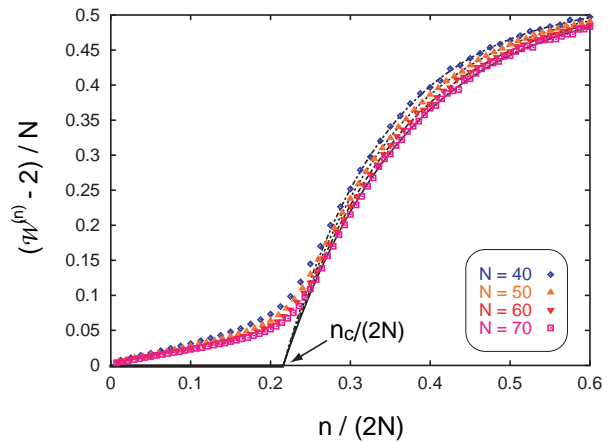}
\end{center}
\caption{(Color online) 
      The normalized Lyapunov localization spectra  
   $\{(\mathcal{W}^{(n)}-2)/N\}_{n}$ as functions of the normalized 
   Lyapunov index $n/(2N)$ in quasi-one-dimensional systems of
   different size; $N=40$ (diamonds), 
   $N=50$ (triangles), $N=60$ (inverted triangles), 
   and $N=70$ (squares) for a density given by $d=10000$.
      The lines are the asymptotic forms of Lyapunov localization 
   spectra in the low density limit for these different values of $N$. 
   }
\label{fig4CriValN}\end{figure}  

   To draw the solid line shown in Fig. \ref{fig1CriValDensi} 
we have already used the value obtained in Eq. (\ref{CritiValue}). 
   The parameters $\alpha$ and $\beta$ in Eq. 
(\ref{LowDensityLimitLine}) are regarded as 
fitting parameters, and we used values of 
$(\alpha,\beta) \approx (2.51,7.26)$ for the line 
in Fig. \ref{fig1CriValDensi}.  
%
%
   The result shown in Fig. \ref{fig1CriValDensi},  
using the critical value (\ref{CritiValue}) 
gives a very satisfactory fit.

   In order to check that the result (\ref{CritiValue}) is 
satisfied for any number of particles $N$, we present 
Fig. \ref{fig4CriValN}.  
   It is the $[n/(2N)]$-dependence of $[\mathcal{W}^{(n)}-2]/N$ 
in quasi-one-dimensional systems of different sizes 
at the same density.
   Notice that values of $[\mathcal{W}^{(n)}-2]/N$ 
themselves decrease slightly as the number of particles increases,    
meaning that the fitting parameters $\alpha$ and $\beta$ in Eq. 
(\ref{LowDensityLimitLine}) depend only slightly on $N$.  
   The values of the fitting 
parameters $\alpha$ and $\beta$ for 
Fig. \ref{fig4CriValN}, were  
$(\alpha,\beta)\approx(2.83,7.83)$ for $N=40$ (dotted broken line), 
$(\alpha,\beta)\approx(2.51,7.26)$ for $N=50$ (dotted line), 
$(\alpha,\beta)\approx(2.28,6.80)$ for $N=60$ (broken line), and 
$(\alpha,\beta)\approx(2.10,6.39)$ for $N=70$ (solid line). 
   However, all the data for different numbers of particles 
in Fig. \ref{fig4CriValN} are nicely fitted using the 
same critical value $n_{c}/(2N)$, given by Eq. (\ref{CritiValue}).

 
   To conclude, we have shown that a model of 
randomly distributed bricks on a line (Fig. \ref{fig2BrickPack}) 
can predict the maximum number of Lyapunov vectors 
which have a Lyapunov localization equal to two, in the low
density limit.
   These are the Lyapunov vectors for which $n<n_{c}$ in the 
linear region
that lead to the bending behavior of Lyapunov localization spectra. 
   We showed that this behavior comes from a restriction 
on the maximum number of the most localized Lyapunov 
vectors with non-zero components for only two particles. 
   The randomly distributed brick model was applied to 
quasi-one-dimensional systems, giving 
the specific value $n_{c}/(2N) \approx 0.216$ for the 
critical value $n_{c}$ of the Lyapunov localization spectra for any 
number of particles $N$. 
   Numerical simulations give a value of $n_{c}/(2N) \approx 0.22 \pm 0.06$. 
   Our explanation for the bending behavior is independent of 
the system width $L_{y}$, as long as the particle order is invariant 
and the system remains quasi-one-dimensional. 
   We have checked this using $L_{y}=2R(2-10^{-6})$ and the
critical value is unchanged.

   The critical value of Lyapunov localization spectra depends 
upon the shape of the system.  
   Numerical results  
show that the critical value of the Lyapunov localization 
spectra for a square system is smaller than that 
for the quasi-one-dimensional system \cite{Tan03b}. 
   The randomly distributed brick model is specific 
to the quasi-one-dimensional system and would need to be
generalized for a square system, and property \condiIIb$\,$ 
can no longer  be assumed. 
   The calculation of  the critical value $n_{c}$ for a  general 
two (or three) dimensional systems remains as a future problem.



   The authors appreciate the financial 
support by the Japan Society for the Promotion Science.




\begin{thebibliography}{99}
\bibitem{note1} In the normalized form 
$\{\mathcal{W}^{(n)}/N\}_n$ of Lyapunov localization 
spectrum, 
its minimum value $2/N$ depends on $N$. 
   To effectively compare Lyapunov localization spectra with 
different numbers of particles, we use the 
normalized Lyapunov localization 
spectra minus $2/N$ 
in Figs. \ref{fig1CriValDensi} and \ref{fig2BrickPack}.   
   This is crucial in Fig. \ref{fig2BrickPack} 
where the particle number dependence of Lyapunov 
localization spectra is discussed.
\end{thebibliography}
\end{document}